\documentclass[aps,prl,twocolumn,groupedaddress]{revtex4}
\usepackage{graphicx}
\usepackage{amsmath}
\usepackage[sort&compress]{natbib}

\newcommand{\rvec}{{\bf r}}      
\newcommand{\cvec}{{\bf c}}     
\newcommand{\tvec}{{\bf t}}      
\newcommand{\lp}{{\ell_p}} 

\begin{document}

\title{Spontaneous Unknotting of a Polymer Confined in a Nanochannel}

\author{Wolfram M\"{o}bius$^{1,2}$}
\author{Erwin Frey$^{2}$}
\author{Ulrich Gerland$^{1}$}

\affiliation{$^{1}$Institute for Theoretical Physics, University of Cologne, Z\"ulpicher Str. 77, 50937 K\"oln, Germany\\
$^{2}$Arnold Sommerfeld Center for Theoretical Physics (ASC) and Center for NanoScience (CeNS), LMU M\"unchen, Theresienstra{\ss}e 37, 80333 M\"unchen, Germany}

\date{\today}

\begin{abstract}
We study the dynamics of a knot in a semiflexible polymer confined to a narrow channel of width comparable to the polymers' persistence length. Using a combination of Brownian dynamics simulations and a coarse-grained stochastic model, we characterize the coupled dynamics of knot size variation and knot diffusion along the polymer, which ultimately leads to spontaneous unknotting. We find that the knot grows to macroscopic size before disappearing. Interestingly, an external force applied to the ends of the confined polymer speeds up spontaneous unknotting.
\end{abstract}


\maketitle

\noindent With recent nanotechnology, single biopolymers can be manipulated and observed inside fabricated nanoscale devices \cite{Austin_NatNanotech_07}. Such `lab-on-a-chip' techniques are interesting to study the physical or biochemical properties of individual molecules, and to sort them according to these properties. A key benefit of confining, e.g., a long DNA molecule in a nanochannel is the resulting stretched-out conformation which permits the mapping of a position along the channel axis to a position on the polymer contour. Thereby, confinement offers the possibility to directly observe and map local sequence-dependent properties of DNA such as affinity for protein binding, which can conventionally be studied only indirectly with biochemical methods. The physics of polymers in confined geometries \cite{Odijk_Macromolecules_83, Reisner_PRL_05, Odijk_JChemPhys_06, Wagner_PRE_07} is fundamental to these applications. An interesting problem arises from topological defects in the form of knots in the polymer contour \cite{Arai_Nature_99, Bao_PRL_03}, which destroy the mapping between channel and polymer coordinates. Knots easily form in a linear polymer, before or while it is threaded into a nanochannel \cite{Metzler_JCompTheorNanos_07}. 

Here, we consider the situation after the polymer is trapped inside the channel. Since a knot in a linear polymer is not topologically conserved, it eventually disappears spontaneously, driven by thermal fluctuations \cite{Bao_PRL_03, Vologodskii_BiophysJ_06, Metzler_EPL_06, Grosberg_PRL_07}. Our focus is on the physics of this process under strong confinement. The typical conformation of a linear polymer with contour length $L$ inside a channel strongly depends on the channel width $d$. It forms a random coil like a free polymer, if the channel is wider than the polymers' radius of gyration (and filled with a good solvent). In more narrow channels, the conformation elongates as described within the blob picture \cite{deGennes_book}. Further reducing $d$ to the scale of the polymers' persistence length $\lp$ gradually stretches out the polymer, by constraining the local tangent vectors $\tvec(s)$ of its contour ($0<s<L$) to directions close to the direction $\cvec$ of the channel axis \cite{Wagner_PRE_07}. Concomitantly, the typical number of ``U-turns'', i.e., sign reversals of $\tvec(s)\cdot\cvec$, decreases to zero \cite{Odijk_JChemPhys_06}. Recent experiments have entered this strong confinement regime with DNA molecules \cite{Reisner_PRL_05}. In this regime, the characteristic lengthscale of the polymer conformation is the Odijk length $L_d\sim(d^2\lp)^{1/3}$, the typical distance between subsequent collisions of the polymer contour with the channel wall \cite{Odijk_Macromolecules_83}. 

\begin{figure}[b]
\includegraphics[width=1.0\columnwidth]{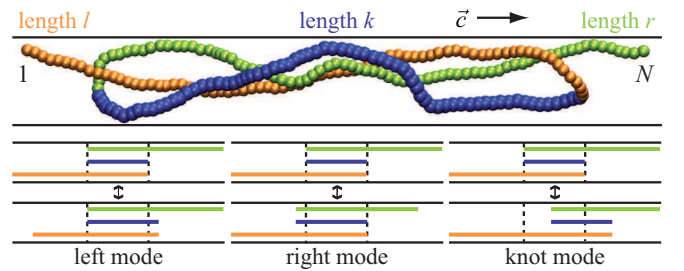}
\caption{\label{fig:knotsegments}
(Top) Illustration of a knotted polymer confined to a nanochannel, and definition of the left, center, and right segment with lengths $l$, $k$, and $r$, respectively. (Bottom) Illustration of the three slow modes of motion allowing the knot to move and change its size.}
\end{figure}

In a knotted configuration, at least two U-turns are topologically interlocked, see Fig.~\ref{fig:knotsegments} (top). These U-turns can disappear only at the polymer ends (their ``pair annihilation'' is impossible), either at the same or at different ends. In terms of the knot, these two options correspond to spontaneous unknotting via diffusion of the knot to one end, or to swelling of the knot to macroscopic sizes of order $L$. Knot diffusion was experimentally observed in DNA molecules stretched by an externally applied tension \cite{Bao_PRL_03}, and addressed in recent theoretical studies \cite{Vologodskii_BiophysJ_06, Metzler_EPL_06, Grosberg_PRL_07}. However, the dynamics of confined knots was either not considered \cite{Vologodskii_BiophysJ_06, Grosberg_PRL_07} or treated on an equal footing with force-induced stretching \cite{Metzler_EPL_06}. We find that these two situations lead to qualitatively different dynamics. In particular, we identify a dynamic interplay between knot diffusion and swelling under strong confinement, which is absent for force-induced stretching.

Due to the enormous range of timescales in the system, we take a multiscale approach to characterize its dynamics. The shortest relevant timescales are independent of $L$ and correspond to the local relaxation of the polymer degrees of freedom in, e.g., a single Odijk length, whereas the longest relevant timescale is the lifetime of a knot, which increases strongly with the length of the polymer, see below. We use Brownian dynamics simulations to explore the full polymer dynamics over short periods, and then match the results to a coarse-grained stochastic model for the long term dynamics. 

To study the full polymer dynamics, we use a `bead-spring' polymer model consisting of $N$ beads at positions $\rvec_{i}$ coupled by a harmonic spring potential, $U_s=K_s\sum_i (|\rvec_{i,i+1}|-b)^{2}/2$ with $\rvec_{i,j}=\rvec_{i}-\rvec_{j}$, which keeps neighboring beads approximately at a distance $b$. A soft excluded volume potential, $U_{ex} = 3 k_BT \sum_{j<i-1} g(\sigma/|\rvec_{i,j}|)$ prevents self-crossing of the polymer. Here, $\sigma$ denotes the interaction range, and $g(x)=(x^{12}\!-\!2x^{6}\!+\!1)\Theta(1-x)$ yields the repulsive part of the Lennard-Jones potential with the Heaviside function $\Theta(x)$. 
The bending stiffness is described by the discretized worm-like chain energy $U_b=k_BT\lp/b\cdot\sum_i(1-\cos\theta_i)$ with the local bending angle $\theta_i$ at bead $i$. The polymer is confined to a channel with circular profile by a soft wall potential,
$U_{ch} = k_BT \epsilon^{-4} \sum_i \left(h_i-\frac{d}{2}+\epsilon\right)^4 \Theta(h_i - \frac{d}{2} + \epsilon)$, where $h_i$ is the distance of bead $i$ from the channel axis, $d$ the channel diameter, and $\epsilon$ parameterizes the softness of the potential. We measure all lengths in units of $d$ and energies in units of $k_BT$, i.e., $d=1$ and $k_BT=1$ in numerical simulations. To put us in the strong confinement regime, we choose a persistence length $\lp=3$ and consider only polymer lengths $L=Nb\gg \lp$. Furthermore, we use a spring constant $K_s=10^4$, a segment length $b=0.1$, a bead-bead interaction range $\sigma=1.2\,b$, and $\epsilon=0.05$. 

Taken together, we have the energy function $U=U_s+U_{ex}+U_b+U_{ch}$. The Brownian dynamics of the polymer is then described by the discrete-time Langevin equations,
\begin{equation}
\rvec_i(t+\Delta t) = \rvec_i(t) -\mu_b\Delta t\,\nabla_{\rvec_i}U + \sqrt{2\mu_b\,k_BT\,\Delta t}\,\textrm{\boldmath{$\eta$}}_{i}(t)
\label{eq:BD}
\end{equation}
with a bead mobility $\mu_b$, a time step $\Delta t$, and random forces {\boldmath{$\eta$}}$_i(t)$ with a variance $\langle\textrm{\boldmath{$\eta$}}_{i}(t)\cdot\textrm{\boldmath{$\eta$}}_{j}(t')\rangle=3\,\delta_{ij}\delta_{tt'}$. 
We choose our time unit such that a polymer segment of unit length has unit mobility, $\mu_b=b$, and a time step $\Delta t=2\cdot10^{-7}$, a tradeoff between accuracy and efficiency. 

Eq.~(\ref{eq:BD}) describes the coupled dynamics of the beads, but we are ultimately interested in how the conformation evolves on a coarse-grained scale. To this end, we consider simple trefoil knots as in Fig.~\ref{fig:knotsegments}, and monitor two observables, the knot size $k$ and position $p$ defined as 
\begin{equation}
k \equiv b\sum_i \Theta(\cvec\cdot\rvec_{i,i+1}) \;,\quad 
p \equiv \frac{b^2}{k} \sum_i i\,\Theta(\cvec\cdot\rvec_{i,i+1}) \;.
\label{eq:CGcoordinates}
\end{equation} 
Here $\cvec$ is a vector along the channel axis, with the same orientation as $\rvec_{N,1}$, see Fig.~\ref{fig:knotsegments}. These definitions are applicable in the strong confinement regime, where spontaneous formation of U-turns is suppressed and a trefoil knot has only two interlocked U-turns. The length $k$ of the polymer contour between the turns is then a well-defined measure of the knot size, as is the location $p$ of its midpoint for the knot position.

\begin{figure}[t]
\includegraphics[width=1.0\columnwidth]{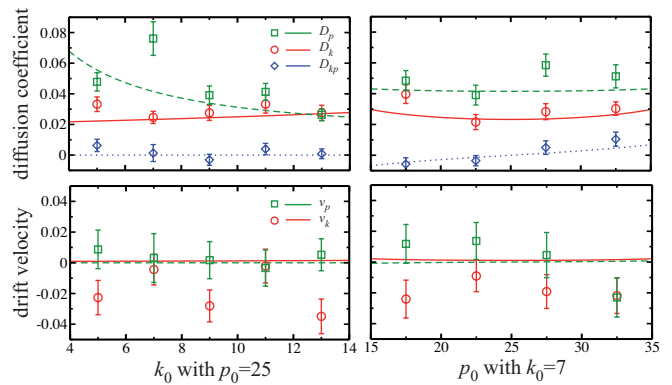}
\caption{Diffusion coefficients ($D_p$, $D_k$, $D_{kp}$, top) and drift velocities ($v_p,v_k$, bottom) for various initial $k_0$ (left) and $p_0$ (right) obtained from Brownian dynamics simulations (data points) \cite{note:bootstrapping} and a coarse-grained model (lines). 
\label{fig:knotpossizemeanvariance}}
\end{figure}

We use initial polymer configurations drawn from an ensemble with a specified knot position $p_0$ and size $k_0$ \cite{note:equilibration}, and evolve them according to Eq.~(\ref{eq:BD}). We find that the knot diffuses along the polymer and at the same time changes its size, apparently in an erratic, non-biased way. To quantify this behavior, we consider the time-dependent position shift $\delta p(t) \!=\! p(t)-p_0$ and size shift $\delta k(t) \!=\! k(t)-k_0$, and determine their averages, variances, and covariance from $150$ independent simulations for each combination of initial values $k_0$ and $p_0$. For these simulations, we used total simulation times considerably longer than required for relaxation of the nearly straight polymer sections, but short enough to avoid large changes in $p$ or $k$ within a simulation. The data is thus suitable to estimate local drift velocities $v_{p}$, $v_{k}$ and diffusion coefficients $D_{p}$, $D_{k}$ in knot position and size space. The symbols in Fig.~\ref{fig:knotpossizemeanvariance} show these estimates for a polymer of length $L=50$, both as a function of $k_{0}$ at a fixed $p_{0}$ and as a function of $p_{0}$ with fixed $k_{0}$. Also shown is the cross-correlation coefficient $D_{kp}=\partial_{t}\langle\delta p\,\delta k\rangle/2$. 

The drift/diffusion coefficients $v_{p}$, $v_{k}$, $D_{p}$, $D_{k}$, and $D_{kp}$ provide the ``interface'' between our detailed Brownian dynamics analysis and our coarse-grained description in $(k,p)$-space. In the following we will see that the dependence on $k_{0}$ and $p_{0}$ observed in Fig.~\ref{fig:knotpossizemeanvariance} is indeed very plausible. We then study the implications of this dependence for the long term knot dynamics. 

On a larger scale, an equation for the stochastic dynamics of the knot in $(k,p)$-space can be derived if the relevant modes of motion and their associated mobilities are known. 
The long time dynamics is dominated by the slow modes where a large segment of the polymer moves relative to the solvent. For our trefoil knot, there are three segments, the central segment between the two U-turns with contour length $k$, the left end with length $l=p-k/2$, and the right end with length $r=L-p-k/2$. As illustrated in Fig.~\ref{fig:knotsegments}, each of these segments has an associated mode: 
(i) In a ``left mode'', the left end moves and exchanges length with the center segment. 
(ii) Similarly, in a ``right mode'', the right end moves, exchanging length with the center. 
(iii) In contrast, a ``knot mode'' exchanges polymer length between the left and the right end, but only the center segment moves. 
The mobility of each mode scales inversely with the length of the segment moving relative to the solvent. 

The mode approach leads to coupled stochastic differential equations for the knot position $p(t)$ and size $k(t)$ in the general Ito type form 
\begin{equation}
\frac{d}{dt} \left( \begin{array}{c}k \\ p\end{array}\right) = \nabla\,\mathbf{D} -\mathbf{M}\,\nabla\,U(k,p) + \sqrt{2}\,\mathbf{B} \,\textrm{\boldmath{$\eta$}}(t) \;.
\label{eq:LE}
\end{equation}
Here, the diffusion matrix $\mathbf{D}$ is linked to the mobility matrix $\mathbf{M}$ via the Einstein relation $\mathbf{D}=k_BT\,\mathbf{M}$. 
The potential $U(k,p)$ can be used to study the effect of additional forces, see below, but is set to zero for now. 
The noise vector {\boldmath{$\eta$}} has three uncorrelated components $\eta_{i}$, one for each mode, with $\langle\eta_i(t)\rangle=0$ and $\langle\eta_i(t)\eta_j(t')\rangle=\delta_{ij}\delta(t-t')$. The matrix $\mathbf{B}$ describes the effect of the three noise components on the coordinates $(k,p)$. It is connected to the diffusion matrix via $\mathbf{D}=\mathbf{B}\mathbf{B}^\mathrm{T}$. Hence, knowledge of $\mathbf{B}$ suffices to specify the coarse-grained dynamics explicitly. For the above modes, we find 
\begin{equation}
\mathbf{B}=\frac{1}{4} \left(\begin{array}{ccc}
2/\sqrt{l} & 2/\sqrt{r} & 0 \\ \noalign{\medskip}
-1/\sqrt{l} & 1/\sqrt{r} & 2/\sqrt{k}
\end{array}\right)\;,
\label{eq:B}
\end{equation}
where we again set the mobility of a polymer stretch of unit length as well as $k_BT$ equal to one.
The diffusion matrix and its derivatives yield simple analytical expressions, without any adjustable parameters, for our drift/diffusion coefficients. Conversely, the five coefficients fully specify the infinitesimal motion in $(k,p)$-space, and the long term dynamics is obtained by integrating Eq.~(\ref{eq:LE}) or the equivalent Fokker-Planck equation. 
Fig.~\ref{fig:knotpossizemeanvariance} suggests that the three slow modes correctly capture the coarse-grained dynamics of the knot (see, however, the closer analysis below). 

The modes also facilitate the physical interpretation of Fig.~\ref{fig:knotpossizemeanvariance}. For instance, the knot mode dominates the mobility of the knot, $D_{p}=\frac{1}{16}(\frac{4}{k}+\frac{1}{l}+\frac{1}{r})$, with a contribution that depends inversely on $k$ (smaller knots are more mobile). Knot size fluctuations increase when the knot approaches a polymer end, $D_{k}=\frac{1}{4}(\frac{1}{l}+\frac{1}{r})$, and $D_{k}$ also grows slowly with knot size due to the concomitant shrinking of the ends. The cross-correlation $D_{kp}=\frac{1}{8}(\frac{1}{r}-\frac{1}{l})$ is appreciable only close to the edges, where the dominant end mode simultaneously leads to a shift in the knot position and size. 
Within the mode approach, drift coefficients arise only via the noise-induced drift term, $\nabla\,\mathbf{D}$. The size drift $v_{k}=\frac{1}{4}(\frac{1}{l^2}+\frac{1}{r^2})$ tends to increase the knot size, whereas the positional drift $v_{p}=\frac{1}{8}(\frac{1}{r^2}-\frac{1}{l^2})$ tends to push knots to the edges. 

Since even a small drift velocity dominates over diffusion at long timescales, the knot dynamics is potentially very sensitive to any drift effects. Both drift effects discussed above are very small and our analysis below shows they are not relevant over the typical lifetime of a knot. However, in the simulation data in Fig.~\ref{fig:knotpossizemeanvariance}, the statistical errors for the drift coefficients are comparable to the average values, and both are much larger than the noise-induced drift. The positional drift values $v_{p}(k_0,p_0)$ show no systematic bias and are compatible with zero or just the noise-induced drift. In contrast, the size drift values $v_{k}(k_0,p_0)$ display a bias to negative values, suggesting systematic knot shrinkage. To pinpoint the origin of this effect, we first tested whether it depends on the topology of the polymer configuration: 
We repeated the simulations starting from initial configurations with two U-turns and the same $k_0,p_0$ values, but without knot. These simulations resulted in the same bias as for the knotted configuations (data not shown). 
We then repeated the simulations again, but with the excluded volume potential $U_{ex}$ switched off. In this case, we observed no bias (data not shown). 

These results indicate that the self-exclusion between the three polymer strands in the channel region between the U-turns causes the bias. 
The self-exclusion effect is noticeable in our simulations due to the large bead size, which is required for computational efficiency. However, under typical experimental conditions it would be negligible except under extreme confinement. 
On the theoretical side, we expect no additional drift effects, since the entropy of the three polymer segments of Fig.~\ref{fig:knotsegments} is additive \cite{Odijk_Macromolecules_83}, and the bending energy of the U-turns does not depend on $k$. Note that the confinement is essential for these arguments; knots in unconfined semiflexible polymers display different physics \cite{Grosberg_PRL_07}. 

\begin{figure}[b]
\includegraphics[width=1.0\columnwidth]{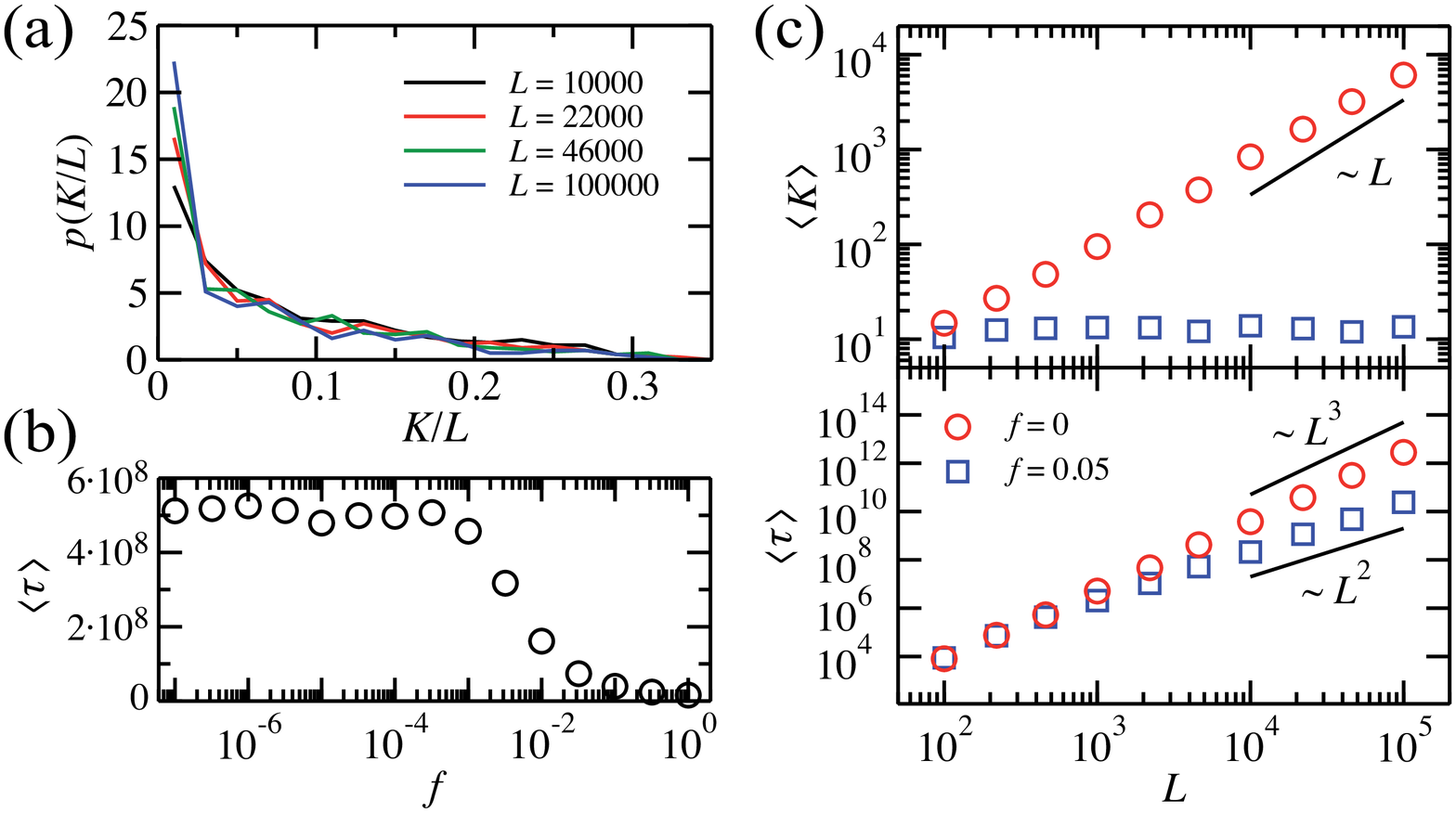}
\caption{(a) Distribution of final knot sizes $K$ at unknotting, scaled by the polymer length $L$. The broad distribution shows that knots become macroscopically large. 
(b) Effect of an applied external force $f$ on the mean knot lifetime $\langle\tau\rangle$ for $L=5000$. Above a threshold the $\langle\tau\rangle$ becomes much smaller, i.e., unknotting is accelerated. 
(c) Scaling of mean final knot size $\left<K\right>$ and mean knot lifetime $\left<\tau\right>$ with polymer length $L$ for $f=0$ and $f=0.05$. 
\label{fig:CGModel}}
\end{figure}

Next, we explore the long term knot dynamics. Since changes in $k$ and $p$ occur on similar timescales, see Fig.~\ref{fig:knotpossizemeanvariance}, we expect a dynamic interplay between these two degrees of freedom. Specifically, we address the question whether knots disappear by growing, by diffusing to the end, or by a combination of both processes. 
To this end, we employ the coarse-grained eq.~(\ref{eq:LE}) and incorporate the physical constraint that the knot cannot shrink to arbitrarily small size via the potential $U(k,p)$ with a steeply rising barrier at $k=1$. A natural definition of knot disappearance is ${\mathrm{min}}(l,r)<k$. Starting with a small knot ($k_0=5$) in the center of the polymer ($p_0=L/2$), we determine the final knot size $K$ just before unknotting. Fig.~\ref{fig:CGModel}(a) shows the resulting distribution of $K/L$ for different polymer lengths $L$. 
The broad range of final knot sizes indicates that polymers neither unknot by knot growth nor by knot diffusion alone, but by a combination of both processes, and that knots typically grow to macroscopic sizes $\sim L$ before disappearing. 

This scaling behavior can be rationalized with a simple argument. Consider a knot diffusing along the polymer until it dissolves after a time $\tau$. It reaches a typical size $k\sim\sqrt{D_k \tau}$ with $D_k$ on the order of $1/L$. Hence, $k^2 \sim\tau/L$. On the other hand, the typical time until the knot reaches a polymer end scales as $\tau\sim L^2/D_p$ with $D_p\sim 1/k$. Taken together, the knot has a characteristic final size $k\sim L$ and lifetime $\tau \sim L^3$. By the same arguments, one finds that the noise-induced drift is (marginally) irrelevant, i.e., it does not affect the scaling of $\tau$. Generally, any drift effect on the knot size is significant only when $v_{k}^{2}>D_{k}/\tau$. 

How is the knot dynamics affected when a weak external pulling force $f$ is applied to the ends of the confined polymer? We consider the force regime where the knot size is constrained without tightening the knot so that molecular friction would become important. 
We include the pulling force through an additional term $f\!\cdot\!(r\!+\!l\!-\!k)=f\!\cdot\!(2k\!-\!L)$ in the potential $U(k,p)$. Fig.~\ref{fig:CGModel}(b) shows the mean unknotting time $\langle\tau\rangle$ as a function of the applied force. While the unknotting time is independent of the applied force for small $f$, it then {\em decreases}, i.e., an increase in the tension accelerates unknotting. 
This result becomes intuitively clear when considering that forces much larger than $\sim k_BT/L$ constrain the knot to sizes much smaller than the polymer length; the knot then mainly diffuses along the polymer with constant mobility, leading to a typical unknotting time $\tau \sim L^2$, faster than the $L^3$-scaling without force, see Fig.~\ref{fig:CGModel}(c). 
For experimentally relevant parameters, the cross-over of Fig.~\ref{fig:CGModel}(b) occurs at very small forces, e.g., in the sub-pN regime for a DNA of $1\,\mu\textrm{m}$ length. 

To summarize, we showed that knot dynamics under strong confinement is qualitatively different from knot dynamics under tension. While the polymer is stretched out in both cases, the knot remains localized in the latter, whereas in the former, knot sizes and knot position change simultaneously, leading to knot growth up to macroscopic sizes.
As a consequence, tiny forces which keep the knot localized can speed up unknotting considerably. Experimentally, such forces can easily occur, e.g., by weak attachment and hydrodynamic drag.

Our theoretical treatment can be generalized to more complicated knots, which may be associated with more U-turns and associated modes. However, our model is clearly very simplistic in several respects. Most importantly, we did not explicitly consider hydrodynamic and electrostatic interactions between the wall and the polymer, as well as between different parts of the polymer. These effects certainly influence the absolute timescale of the knot dynamics. However, we expect that they do not affect the dynamics qualitatively. For instance, electrostatic effects should be weak as long as the Debye screening length is considerably smaller than the channel diameter \cite{Dommersnes_PRE_02}. We hope that our assumptions and predictions will be scrutinized experimentally and help to explore the physics of polymers in nanofluidic devices. 

Fruitful discussions with Walter Reisner, Richard Neher, and Ralf Metzler, as well as financial support by the DFG via SFB 486 and the German Excellence Initiative via the program NIM are gratefully acknowledged.


\vspace{-0.5cm}

\begin{thebibliography}{99}

\vspace{-0.5cm}

\bibitem{Austin_NatNanotech_07}
R.~Austin, Nat. Nanotech. {\bf 2}, 79 (2007). 

\bibitem{Odijk_Macromolecules_83}
T.~Odijk, Macromolecules {\bf 16}, 1340 (1983).

\bibitem{Reisner_PRL_05}
W.~Reisner {\it et al.}, Phys. Rev. Lett. {\bf 94}, 196101 (2005).

\bibitem{Odijk_JChemPhys_06}
T.~Odijk, J. Chem. Phys. {\bf 125}, 204904 (2006).

\bibitem{Wagner_PRE_07}
F.~Wagner, G.~Lattanzi, and E.~Frey, Phys. Rev. E {\bf 75}, 050902(R) (2007).

\bibitem{Arai_Nature_99}
Y.~Arai, Nature {\bf 399}, 446 (1999).

\bibitem{Bao_PRL_03}
X.~R.~Bao, H.~J.~Lee, and S.~R.~Quake, Phys. Rev. Lett. {\bf 91}, 265506 (2003); Phys. Rev. Lett. {\bf 95}, 199901 (2005).

\bibitem{Metzler_JCompTheorNanos_07}
R.~Metzler {\it et al.}, J. Comput. Theor. Nanos. {\bf 4}, 1 (2007).

\bibitem{Metzler_EPL_06}
R.~Metzler {\it et al.}, Europhys. Lett. {\bf 76}, 696 (2006).

\bibitem{Vologodskii_BiophysJ_06}
A.~Vologodskii, Biophys. J. {\bf 90}, 1594 (2006).

\bibitem{Grosberg_PRL_07}
A.~Y.~Grosberg and Y.~Rabin, Phys. Rev. Lett. {\bf 99}, 217801 (2007).

\bibitem{deGennes_book}
P.-G.~de Gennes, {\it Scaling Concepts in Polymer Physics} (Cornell University Press, 1979).

\bibitem{Dommersnes_PRE_02}
P.~G.~Dommersnes, Y.~Kantor, and M.~Kardar, Phys. Rev. E {\bf 66}, 031802 (2002).

\bibitem{note:knotdef}
To avoid effects by fluctuating ends, the sums do not contain the beads at the very ends of the polymer.

\bibitem{note:equilibration}
To generate an ensemble of polymer configurations with knot size $k_0$ and position $p_0$, we equilibrated the Brownian dynamics simulation in an additional external potential $U_{IC}=K_k (k-k_0)^2/2+K_p (p-p_0)^2/2$ with $k$, $p$ expressed in terms of the bead coordinates (see main text) and $K_k \!=\! K_p \!=\! 1000$. 

\bibitem{note:bootstrapping}
For each ($k_0,p_0$) 150 Brownian dynamics trajectories were simulated for $0<t<5$ following equilibration. $D_p$, $D_k$, $D_{pk}$, $v_p$, and $v_k$ (for definition see main text) were determined by fits on 5000 different ensembles of $\langle\delta k\rangle(t)$ and $\langle\delta p\rangle(t)$ drawn from the 150 trajectories with replacement in a bootstrapping fashion. Fits were done for $t>1$ to allow for microscopic rearrangements following equilibration. For the same reasons, ordinate intercept was allowed when deducing $D_k$, $D_p$, and $D_{kp}$.

\end{thebibliography}
\end{document}